\documentclass[twocolumn,showpacs,preprintnumbers,amsmath,amssymb, unsortedaddress,floatfix]{revtex4}
\usepackage{amsfonts}
\usepackage{mathrsfs}

\usepackage{amsmath}
\usepackage{graphicx}
\usepackage{bm}
\usepackage{subfigure}

\newcommand{\ket}[1]{\mbox{$|#1\rangle$}}

\begin{document}

\title{Dependence of the decoherence of polarization states in phase-damping channels on the frequency spectrum envelope of photons}

\author{Yan-Xiao Gong}
\email{yxgong@mail.ustc.edu.cn}
\author{Yong-Sheng Zhang}
\email{yshzhang@ustc.edu.cn}
\author{Yu-Li Dong}
\author{Xiao-Ling Niu}
\author{Yun-Feng Huang}
\email{hyf@ustc.edu.cn}
\author{Guang-Can Guo}
\affiliation{Key Laboratory of Quantum Information, University of
Science and Technology of China, CAS, Hefei, 230026, People's
Republic of China}

\begin{abstract}

We consider the decoherence of photons suffering in phase-damping
channels. By exploring the evolutions of single-photon polarization
states and two-photon polarization-entangled states, we find that
different frequency spectrum envelopes of photons induce different
decoherence processes. A white frequency spectrum can lead the
decoherence to an ideal Markovian process. Some color frequency
spectrums can induce asymptotical decoherence, while, some other
color frequency spectrums can make coherence vanish periodically
with variable revival amplitudes. These behaviors result from the
non-Markovian effects on the decoherence process, which may give
rise to a revival of coherence after complete decoherence.

\end{abstract}


\maketitle

\section{Introduction}

Photons have been widely applied in quantum teleportation
\cite{bouwmeester_teleportation}, quantum dense coding
\cite{mattle_dense}, quantum cryptography \cite{gisin_QKD}, and
quantum computing \cite{Kok_computing}. Among these, information can
be encoded in any of the degrees of freedom (DOF) of photons, such
as polarization, frequency, momentum angular, path, or energy time.
If we only consider the quantum state in some of the DOF, the other
DOF can be regarded as ``environment'' \cite{kwiat_DFS,
Berglund_decoherence, altepeter_DFS, white_mixed, Thew_mixed,
zhang_werner, Peters_mixed, Barbieri_mixed, puentes_mixed,
Almeida_death}. In this context, coupling between different DOF
would destroy the coherence of the quantum state considered, leading
to decoherence effects, which limit the practical implementation of
quantum information processing \cite{zurek_decoherence}.

In this paper, we consider a non-dissipative coupling between photon
frequency and polarization in a birefringent media, resulting in
decoherence of polarization due to different group velocities for
two orthogonal polarization modes. Coupling between these two DOF
has been widely studied in many optical experiments, especially in
the field of polarization mode dispersion in optical fibers
\cite{Gordon_PMD, Nelson_PMD}. Here we simply focus on the
phase-damping channels composed of birefringent crystals with fixed
optic axes.  As this decoherence can be easily controlled by
rotating the optic axes or changing the length of birefringent
crystals, it has been widely utilized in the experimental research
on quantum decoherence dynamics of photons, such as verifying
decoherence-free space \cite{kwiat_DFS, Berglund_decoherence,
altepeter_DFS}, characterizing entangled mixed states
\cite{white_mixed, Thew_mixed, zhang_werner, Peters_mixed}.
Therefore, it is intuitive and reasonable  to investigate the
dependence of the decoherence process in this model on the spectrum
envelopes of the ``environment'', namely, the frequency spectrum
envelopes (FSEs) of photons.

The aim of this paper is to address this point by exploring the
behaviors of normalized linear entropy \cite{Bose_entropy} of
single- and two-photon polarization states, and concurrence
\cite{Wootters_concurrence, Coffman_concurrence} of two-photon
polarization states. We find that the behaviors of these quantities
against the length of birefringent crystals depend on the photon
spectrum forms. If the FSE is an ideal white spectrum, the
decoherence is Markovian. However, any color spectrums result in
non-Markovian decoherence processes. In particular, some FSEs lead
to asymptotical decoherence, while, some FSEs induce coherence to
vanish periodically with variable revival amplitudes.

Recently, much interest has arisen in the roles of non-Markovian
effects played on quantum states evolution \cite{nonMar_ahn,
nonMar_lee, nonMar_shresta, nonMar_ban, nonMar_ban2,
nonMar_glendinning, nonMar_maniscalco, nonMar_bellomo, nonMar_liu,
nonMar_dajka, nonMar_piilo, nonMar_bellomo2}. The ``memory effects''
of the non-Markovian reservoir can preserve the coherent information
of the quantum system during its relaxation time. To show this
effect, we obtain the time correlation functions corresponding to
different spectrum functions. The correlation function gets a
delta-function form when the wave-packet takes the white spectrum,
namely the flat spectrum. So in this case, the decoherence is
Markovian. If the wave-packet takes non-flat spectrum, the
correlation function is a non-trivial function of time, and this
behavior can lead a typical non-Markovian effect. Our results then
open the door to experimental investigations of the non-Markovian
decoherence processes. Moreover, our results on the dynamics of
two-photon entanglement, present a possible way to experimental
research on entanglement sudden death \cite{death_yu1, death_yu2,
death_yu3, Almeida_death} and recovering entanglement after complete
disentanglement.

The rest of our paper is organized as follows. In the next section,
we consider the decoherence of single-polarization states and show
the effects of FSEs on the decoherence process by giving several
examples of FSEs. In Sec.~\ref{sec_two}, we investigate the roles of
FSEs played on the evolution of two-photon polarization entangled
states, through behaviors of linear entropy and concurrence with
some examples of FSEs. In Sec.~\ref{sec_conclude}, we conclude and
discuss the experimental feasibility of our results with present
photon sources.

\section{Decoherence of single-photon polarization states} \label{section_single}

We first review the evolution of pure single-photon states in the
phase-damping channel based on the calculations in
Ref.~\cite{Berglund_decoherence}. An arbitrary single-photon state
characterized by its polarization and frequency spectrum can be
represented as \cite{Berglund_decoherence},
\begin{equation}\label{eq_initial}
\ket{\Psi(0)}=(\alpha\ket{H}+\beta\ket{V})\otimes\int d\omega
f(\omega)\ket{\omega},
\end{equation}
where $\ket{H}$ ($\ket{V}$) denotes the horizontal (vertical)
polarization state with arbitrary complex amplitudes $\alpha$ and
$\beta$ satisfying
\begin{equation}\label{eq_normalization1}
|\alpha|^2+|\beta|^2=1,
\end{equation}
and $f(\omega)$ is the complex amplitude corresponding to the
frequency $\omega$, with the normalization condition,
\begin{equation}\label{eq_normalization2}
\int d\omega|f(\omega)|^2=\int d\omega F(\omega)=1,
\end{equation}
where we use the notation \mbox{$F(\omega)=|f(\omega)|^2$}.

Note that for simplicity, polarization and frequency in the initial
state we considered is not entangled. The phase-damping channel in
our model is composed of a birefringent crystal with a fixed optic
axis. Without loss of generality, we set the optic axis in
horizontal direction and assume the horizontally polarized photons
travel faster than vertically polarized photons, i.e.,
$n_{H}<n_{V}$. Here $n_{H}$ ($n_{V}$) is the index of refraction
corresponding to horizontal (vertical) polarization. Then after the
photon is transmitted through a birefringent crystal of length $l$,
the output state can be expressed as \cite{Berglund_decoherence},
\begin{align}\label{eq_output}
\ket{\Psi(l)}=&\alpha\ket{H}\otimes\int d\omega f(\omega)e^{i\omega
n_{H}l/c}\ket{\omega} \nonumber\\
&+\beta\ket{V}\otimes\int d\omega f(\omega)e^{i\omega
n_{V}l/c}\ket{\omega}.
\end{align}
Then we can see that polarization and frequency become entangled. To
obtain the polarization state, we trace over the frequency DOF from
the density matrix of the output state above, resulting in the
output state described as
\begin{equation}\label{eq_matrixsingle}
\rho(l)=
\begin{pmatrix}
|\alpha|^2 &
\alpha\beta^*\mathcal{F}^*(l)\\\alpha^*\beta\mathcal{F}(l) &
|\beta|^2
\end{pmatrix},
\end{equation}
where
\begin{equation}\label{eq_corelation}
\mathcal{F}(l)=\int d\omega F(\omega)e^{i\omega \Delta n l/c},
\end{equation}
and $\Delta n=n_V-n_H$. It should be noted that here for simplicity
we neglect the variation of the refraction index with $\omega$,
since although $n_H$ and $n_V$ depend on frequency $\omega$, the
value of $\Delta n$ does not vary obviously according to $\omega$.

From Eq.~(\ref{eq_corelation}), we can see $\mathcal{F}(l)$ is the
Fourier transform of $F(\omega)$ up to a constant, which depends on
the choice of FSE of the photon. As in our decoherence model, the
crystal length $l$ is proportional to the time $t$, the function of
$\mathcal{F}(l)$ given by Eq.~(\ref{eq_corelation}) represents the
time correlation function in the master equation \cite{noise_book}.
If the function is a non-trivial function of length $l$, the
decoherence is non-Markovian, resulting in non-Markovian effects in
the evolution of quantum states.

To characterize the decoherence of a single-photon state we employ
the normalized linear entropy \cite{Bose_entropy} defined as
\mbox{$S_L\left(\rho\right)\equiv
2\left[1-\text{Tr}\left(\rho^2\right)\right]$} (for pure states
$S_L=0$ and for mixed states $0<S_L\leqslant1$). Then for the output
polarization state the linear entropy is given by
\begin{equation}\label{eq_entropy1}
S_L(l)=2\left[1-\text{Tr}\left(\rho^2\right)\right]=4\left|\alpha\right|^2|\beta|^2\left[1-\left|\mathcal{F}(l)\right|^2\right].
\end{equation}
We can see that the function of linear entropy versus length of the
birefringent crystal depends on the FSE. It is necessary to note
that if $\alpha=0$ or $\beta=0$, $S_L(l)=0$ independent of $l$,
i.e., decoherence does not occur. That is because we set the optic
axis horizontal so that the horizontal and vertical polarization
states are not affected in such a phase-damping channel.

From the density matrix form given by Eq.~(\ref{eq_matrixsingle}),
we can see that the behavior of the linear entropy is equivalent to
that of coherence. In the following part we consider some choices of
FSE and present the behaviors of the linear entropy against the
crystal length to show the variation of coherence.

\subsection{White spectrum}\label{sec_single_white}

We first consider the FSE to be an ideal white spectrum, where
$F(\omega)$ does not depend on the frequency. In this case, it is
meaningless to consider the normalization condition given by
Eq.~(\ref{eq_normalization2}). Usually, the correlation function is
assumed to take the form of a delta-function \cite{noise_book,
white_prataviera}, i.e., \mbox{$\mathcal{F}(l)\sim\delta(l)$}.
Therefore, complete decoherence occurs for any nonzero $l$. In this
context, the decoherence is an ideal Markovian process.

\subsection{Gaussian spectrum}\label{sec_single_gaussian}

Let's then choose the FSE to be a Gaussian form written as
\begin{equation}\label{eq_gau1}
F(\omega)=\frac{1}{\Delta\omega\sqrt{\pi}}\exp\left[-\left(\frac{\omega-\omega_0}{\Delta
\omega}\right)^2\right],
\end{equation}
where $\omega_0$ is the central frequency and $\Delta \omega$
indicates the width of the Gaussian envelope. With this FSE, the
correlation function is found to be
\begin{equation}
\mathcal{F}(l)=\exp\left[-\left(\frac{\Delta n\Delta \omega l}{2
c}\right)^2\right]\exp\left(\frac{i\omega_0\Delta n l}{c}\right),
\end{equation}
and therefore, the linear entropy becomes
\begin{equation}
S_L(l)=4|\alpha|^2|\beta|^2\left\{1-\exp\left[-\frac{1}{2}\left(\frac{\Delta
n\Delta \omega l}{c}\right)^2\right]\right\}.
\end{equation}
We can see that $S_L$ increases Gaussianly and approaches to maximum
asymptotically. Combining the form of the density matrix given by
Eq.~(\ref{eq_matrixsingle}), we can conclude that coherence vanishes
asymptotically.

\subsection{Lorentzian frequency spectrum}

We then consider a Lorentzian frequency spectrum given by
\begin{equation}\label{eq_lor1}
F(\omega)=\frac{\Delta \omega}{\pi}\frac{1}{(\Delta
\omega)^2+(\omega-\omega_0)^2},
\end{equation}
where $\omega_0$ is the central frequency and $\Delta \omega$
denotes the width of the Lorentzian envelope. Then the correlation
function can be expressed as
\begin{equation}
\mathcal{F}(l)=e^{-\Delta n \Delta \omega l/c}e^{i\Delta
n\omega_0l/c}.
\end{equation}
The linear entropy is therefore
\begin{equation}
S_L(l)=4|\alpha|^2|\beta|^2\left(1-e^{-2\Delta n\Delta \omega
l/c}\right).
\end{equation}
It is clear that coherence decays exponentially and vanishes
asymptotically.

\subsection{Rectangular spectrum}

A rectangular spectrum is assumed to take the form
\begin{equation}\label{eq_rec1}
F(\omega)=
\begin{cases}\frac{1}{2\Delta \omega} & |\omega-\omega_0|\leqslant\Delta\omega\\ 0 &|\omega-\omega_0|>\Delta\omega
\end{cases},
\end{equation}
where $\omega_0$ is the central frequency and $\Delta\omega$
represents the width of the rectangular envelope. The correlation
function becomes
\begin{equation}
\mathcal{F}(l)=\text{sinc}\left(\frac{\Delta n\Delta \omega
l}{c}\right)e^{i\Delta n\omega_0 l/c},
\end{equation}
where the function \mbox{$\text{sinc}x\equiv \sin x/x$}. Therefore,
the linear entropy can be expressed as
\begin{equation}
S_L(l)=4|\alpha|^2|\beta|^2\left[1-\left|\text{sinc}\left(\frac{\Delta
n\Delta \omega l}{c}\right)\right|^2\right].
\end{equation}
From the above equation and the behavior of the $\text{sinc}x$, we
can infer that $S_L$ gets to maximum periodically with a damping
decrease. This behavior shows a different decoherence process, i.e.,
coherence vanishes periodically with a damping of its revival
amplitude.

\subsection{Multi-peaked spectrum}

Let's first consider an ideal multi-peaked spectrum, called as
multi-delta spectrum, described by
\begin{equation}\label{eq_doudelta}
F(\omega)=\frac{1}{N}\sum_{j=1}^N\delta(\omega-\omega_j),
\end{equation}
where $\omega_j$ are the peak frequencies. We can see that each peak
is a delta spectrum and that if \mbox{$N\rightarrow \infty $} it is
simply the comb spectrum. With this FSE, the correlation function is
found to be
\begin{equation}
\mathcal{F}(l)=\frac{1}{N}\sum_{j=1}^Ne^{i\Delta n\omega_jl/c},
\end{equation}
and the linear entropy is expressed as
\begin{equation}
S_L(l)=4|\alpha|^2|\beta|^2\left[1-\frac{1}{N^2}\left|\sum_{j=1}^Ne^{i\Delta
n\omega_jl/c}\right|^2\right].
\end{equation}
From the above equation, it is not difficult to infer that $S_L$ can
oscillate between $0$ and $4|\alpha|^2|\beta|^2$ (there would be
some sub-maximum if \mbox{$N>2$}). For simplify, let's analyze the
case of $N=2$, and therefore, the linear entropy becomes
\begin{equation}
S_L(l)=4|\alpha|^2|\beta|^2\left[1-\cos^2\frac{\Delta
n(\omega_1-\omega_2)l}{2c}\right].
\end{equation}
The above function shows that $S_L$ oscillates between $0$ and its
maximum. This behavior indicates that coherence can vanish and
revival periodically.

Actually, in practical experiments, there is no such ideal spectrum,
and the spectrum form at each peak is usually Gaussian, or
Lorentzian, or others. Let's first take the Gaussian form for an
example. For simplicity, we again assume that there are two peaks,
each of which is Gaussian with the same width. The spectrum (called
as double-Gaussian spectrum) is written as
\begin{align}\label{eq_dougau1}
F(\omega)=\frac{1}{2\Delta\omega\sqrt{\pi}}&\left\{\exp\left[-\left(\frac{\omega-\omega_1}{\Delta
\omega}\right)^2\right]\right.\nonumber\\
&+\left.\exp\left[-\left(\frac{\omega-\omega_2}{\Delta
\omega}\right)^2\right]\right\},
\end{align}
where $\omega_1$ and $\omega_2$ are the two peak frequencies and we
assume \mbox{$|\omega_1-\omega_2|=5 \Delta \omega$} so that the two
peaks can be considered absolutely separated. Then we can obtain the
correlation function as
\begin{align}
\mathcal{F}(l)=&\frac{1}{2}\exp\left[-\left(\frac{\Delta n\Delta
\omega l}{2c}\right)^2\right]\nonumber\\
&\times\left[\exp\left(\frac{i\omega_1\Delta n
l}{c}\right)+\exp\left(\frac{i\omega_2\Delta n l}{c}\right)\right],
\end{align}
and the linear entropy as
\begin{align}
S_L(l)=4|\alpha|^2|\beta|^2\Bigg\{1-&\exp\left[-\frac{1}{2}\left(\frac{\Delta
n\Delta \omega l}{c}\right)^2\right]\nonumber\\
&\ \ \times\cos^2\frac{\Delta n(\omega_1-\omega_2)l}{2c}\Bigg\}.
\end{align}
\begin{figure}[tb]
\centering
\includegraphics[width=7cm]{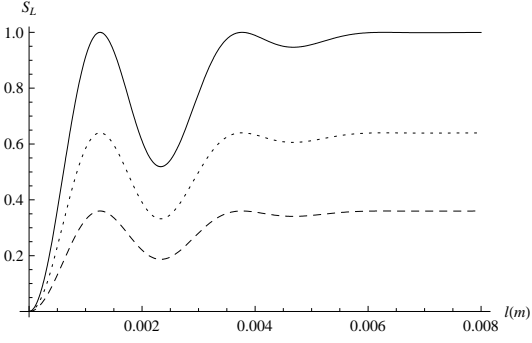}
\caption{Linear entropy $S_L$ for the polarization state
\mbox{$\alpha\ket{H}+\beta\ket{V}$} with a double-Gaussian frequency
spectrum, as a function of crystal length $l$, in the case of
$|\alpha|^2=0.1$ (dashed line), $|\alpha|^2=0.5$ (solid line), and
$|\alpha|^2=0.8$ (dotted line). Here $\ket{H}$ ($\ket{V}$)
represents the horizontal (vertical) polarization state and
\mbox{$|\alpha|^2+|\beta|^2=1$}.} \label{fig_mGau_1s}
\end{figure}
The behaviors of $S_L$ against $l$ are plotted in
Fig.~\ref{fig_mGau_1s}, under the assumption of $\Delta n\Delta
\omega/c=500m^{-1}$, in the case of $|\alpha|^2=0.1$, $0.5$, and
$0.8$. We can see that coherence vanishes periodically with damped
revival amplitudes.

Then we consider an example of double-Lorentzian spectrum given by
\begin{align}\label{eq_doulor1}
F(\omega)=&\frac{\Delta \omega}{2\pi}\bigg[\frac{1}{(\Delta
\omega)^2+(\omega-\omega_1)^2}\nonumber\\
&+\frac{1}{(\Delta \omega)^2+(\omega-\omega_2)^2}\bigg],
\end{align}
where $\omega_1$ and $\omega_2$ are the two peak frequencies and we
assume \mbox{$|\omega_1-\omega_2|=30 \Delta \omega$} so that the two
peaks can be considered absolutely separated. Then the correlation
function can be obtained
\begin{equation}
\mathcal{F}(l)=\frac{1}{2} e^{-\Delta n \Delta \omega
l/c}\left(e^{i\Delta n\omega_1l/c}+e^{i\Delta n\omega_2l/c}\right).
\end{equation}
The linear entropy is therefore
\begin{equation}
S_L(l)=4|\alpha|^2|\beta|^2\left[1-e^{-2\Delta n\Delta \omega
l/c}\cos^2\frac{\Delta n(\omega_1-\omega_2)l}{2c}\right].
\end{equation}
\begin{figure}[tb]
\centering
\includegraphics[width=7cm]{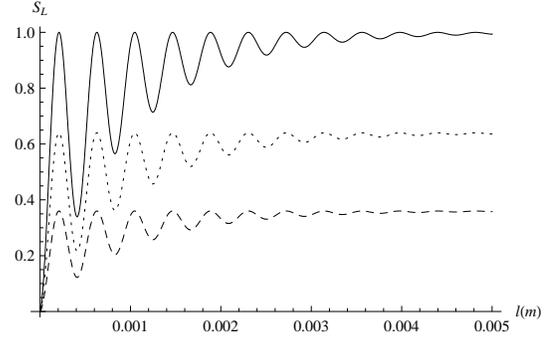}
\caption{Linear entropy $S_L$ for the polarization state
\mbox{$\alpha\ket{H}+\beta\ket{V}$} with a double-Lorentzian
frequency spectrum, as a function of crystal length $l$, in the case
of $|\alpha|^2=0.1$ (dashed line), $|\alpha|^2=0.5$ (solid line),
and $|\alpha|^2=0.8$ (dotted line). Here $\ket{H}$ ($\ket{V}$)
represents the horizontal (vertical) polarization state and
\mbox{$|\alpha|^2+|\beta|^2=1$}.} \label{fig_mLor_1s}
\end{figure}
Fig.~\ref{fig_mLor_1s} shows the behaviors of $S_L$ against $l$,
under the assumption of $\Delta n\Delta \omega/c=500m^{-1}$, in the
case of $|\alpha|^2=0.1$, $0.5$, and $0.8$. We can see that
coherence vanishes periodically with damped revival amplitudes.

\section{Decoherence of two-photon polarization
states}\label{sec_two}

Our analysises above are straightforward to generalize to the
decoherence of arbitrary two-photon polarization states. However,
here we shall only restrict our analysis to the initial two-photon
state given by \cite{Berglund_decoherence}
\begin{align}
\ket{\Phi(0)}=&\left(a\ket{H}_1\ket{H}_2+b\ket{V}_1\ket{V}_2\right)\nonumber\\
&\otimes\int d\omega_1d\omega_2 g(\omega_1,
\omega_2)\ket{\omega_1}_1\ket{\omega_2}_2,
\end{align}
where $a$ and $b$ are arbitrary complex amplitudes satisfying
\begin{equation}\label{eq_normalization3}
|a|^2+|b|^2=1,
\end{equation}
and $g(\omega_1, \omega_2)$ is the complex amplitude corresponding
to frequencies $\omega_1$ and $\omega_2$, with the normalization
condition,
\begin{equation}\label{eq_normalization4}
\int d\omega_1d\omega_2{|g(\omega_1, \omega_2)|^2}=\int
d\omega_1d\omega_2G(\omega_1, \omega_2)=1,
\end{equation}
where the notation \mbox{$G(\omega_1, \omega_2)=|g(\omega_1,
\omega_2)|^2$}. For simplicity, we only consider that only the
photon in mode $1$ is in the phase-damping channel the same with
that in Sec. \ref{section_single} with the other photon in mode $2$
free from decoherence. We should note that analogous analysis can be
applied to the case of both photons suffering in phase-damping
channels. Therefore, after the photon in mode $1$ is transmitted
through a birefringent crystal of length $l$, the state becomes
\cite{Berglund_decoherence}
\begin{align}
\label{eq_output2}
\ket{\Phi(l)}=&a\ket{H}_1\ket{H}_2\nonumber\\
&\otimes\int d\omega_1d\omega_2 g(\omega_1, \omega_2)e^{i\omega_1
n_{H}l/c} \ket{\omega_1}_1\ket{\omega_2}_2\nonumber\\
&+b\ket{V}_1\ket{V}_2\nonumber\\
&\otimes\int d\omega_1d\omega_2 g(\omega_1, \omega_2)e^{i\omega_1
n_{V}l/c}\ket{\omega_1}_1\ket{\omega_2}_2.
\end{align}
Tracing over the frequency DOF from the density matrix of the output
state above, we can get the output polarization state described as
\begin{equation}\label{eq_matrix2}
\rho'(l)=
\begin{pmatrix}
|a|^2 & 0 & 0 & ab^*\mathcal{G}^*(l)\\0 & 0 & 0 & 0\\0 & 0 & 0 &
0\\a^*b\mathcal{G}(l) & 0 & 0 & |b|^2
\end{pmatrix},
\end{equation}
where the correlation function is
\begin{equation}\label{eq_corelation2}
\mathcal{G}(l)=\int d\omega_1d\omega_2 G(\omega_1,
\omega_2)e^{i\omega_1 \Delta n l/c}.
\end{equation}

To quantify the mixedness of the state we use the normalized linear
entropy \cite{Bose_entropy} given by
\begin{equation}\label{eq_entropy2}
S_L(l)=\frac{4}{3}\left[1-\text{Tr}(\rho'^2)\right]=\frac{8}{3}|a|^2|b|^2\left[1-\left|\mathcal{G}(l)\right|^2\right].
\end{equation}
From the density matrix form of Eq.~(\ref{eq_matrix2}), we can see
that the behavior of  $S_L$ can give the variation of coherence.

To measure the entanglement of a two-photon state $\rho$ we employ
the concurrence \cite{Wootters_concurrence, Coffman_concurrence},
given by
\begin{equation}
C(\rho)=\max\left\{0,\sqrt{\lambda_1}-\sqrt{\lambda_2}-\sqrt{\lambda_3}-\sqrt{\lambda_4}\right\},
\end{equation}
where $\lambda_i$ are the eigenvalues of
$\rho(\sigma_y\otimes\sigma_y)\rho^*(\sigma_y\otimes\sigma_y)$, in
nonincreasing order by magnitude and $\sigma_y=\left({0\atop i}
{-i\atop 0}\right)$. The case of $C=0$ means no entanglement between
the two photons and $0<C\leqslant1$ corresponds to the existence of
entanglement between the two photons. Then for the output state
described by Eq.~(\ref{eq_matrix2}), the concurrence is found to be
\begin{equation}\label{eq_con}
C(l)=2|a||b|\left|\mathcal{G}(l)\right|.
\end{equation}

Comparing the Eqs.~(\ref{eq_corelation2}), (\ref{eq_entropy2}) and
(\ref{eq_con}) with Eqs.~(\ref{eq_corelation}) and
(\ref{eq_entropy1}), we can infer that the effects of FSEs on the
decoherence of two-photon states would be similar to that on the
decoherence of single-photon states. It is straightforward to study
the effects for arbitrary forms of $g(\omega_1, \omega_2)$. However,
as the purpose of our paper is to show the effects of different
types of FSE on the evolution of the polarization states rather than
analyze a specific FSE in a practical experiment, it is reasonable
to make the following assumptions.

We restrict our analysis to the two-photon polarization states
generated by parametric down-conversion, which is widely used in
optical experiments. The frequency spectrum takes the form
\cite{shaping_peer}
\begin{align}
g(\omega_1,\omega_2)=h(\omega_1)h(\omega_2)a_p(\omega_p)\varphi(\omega_p,
\omega_1-\omega_2),
\end{align}
with the frequency-anticorrelated relation
\mbox{$\omega_p=\omega_1+\omega_2$}, where $h(\omega_1)$
($h(\omega_2)$) represents the transmission function of the optical
filter, and $a_p(\omega_p)$ describes the pump field spectrum
corresponding to the pump frequency $\omega_p$, and
$\varphi(\omega_p, \omega_1-\omega_2)$ is the phase-matching
function dependent on the size of the non-linear crystal.  If the
non-linear crystal is thin enough, $\varphi(\omega_p,
\omega_1-\omega_2)$ could be neglected. We make a further
simplification by assuming the bandwidth the pump field is very
narrow so that $\omega_p$ could be considered as a constant. We then
have a factorizable spectrum form
\begin{equation}
g(\omega_1,\omega_2)\approx h(\omega_1)h(\omega_2).
\end{equation}

For an ideal white spectrum, again we take the correlation function
as a delta-function, i.e.,\mbox{$\mathcal{G}(l)\sim\delta(l)$}. In
this context, it is clear that both coherence and concurrence vanish
for any nonzero $l$, showing an ideal Markovian decoherence process.

\begin{figure}
\begin{center}
\subfigure[]{
\resizebox*{7cm}{!}{\includegraphics{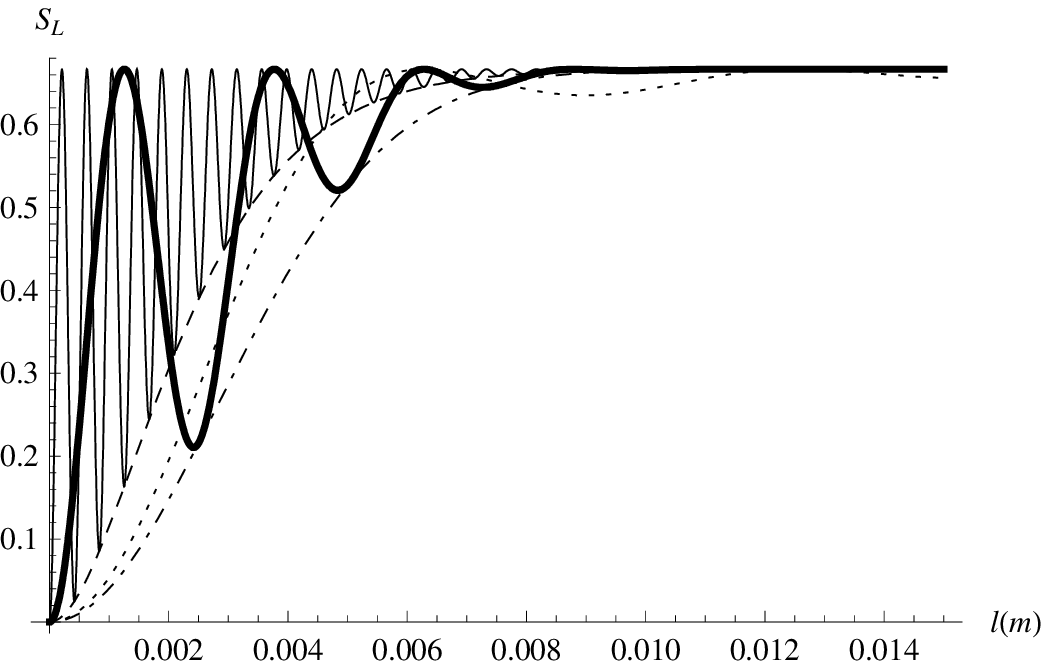}}\label{fig_s2}}
\subfigure[]{
\resizebox*{7cm}{!}{\includegraphics{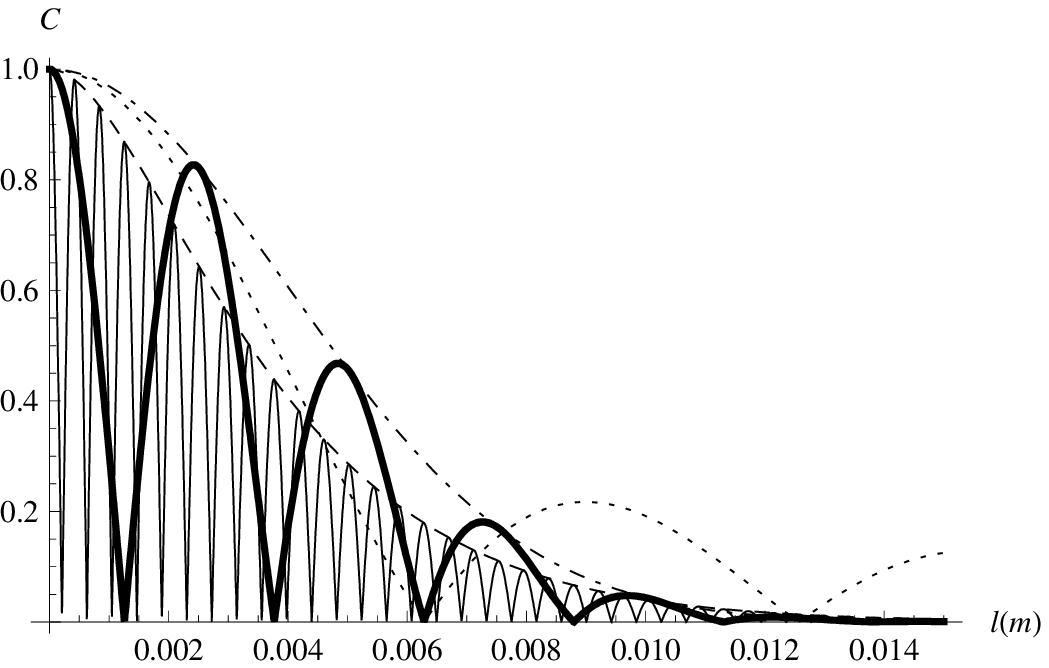}\label{fig_c2}}}
\caption{(a) Linear entropy $S_L(l)$ (b) Concurrence $C(l)$ of the
initial polarization state as a function of crystal length $l$ with
a Gaussian spectrum (dot-dashed line), a Lorentzian spectrum (dashed
line), a rectangular spectrum (dotted line), a double-Gaussian
spectrum (thick solid line) and a double-Lorentzian spectrum (thin
solid line).}
\end{center}
\end{figure}

For color spectrums, by making \mbox{$|h(\omega_1)|^2$}
\mbox{($|h(\omega_2)|^2$)} take the forms given by
Eqs.~(\ref{eq_gau1}), (\ref{eq_lor1}), (\ref{eq_rec1}),
(\ref{eq_dougau1}) and (\ref{eq_doulor1}) (normalization constants
may be needed to satisfy Eq.~(\ref{eq_normalization4})), we give
some
examples as follows,\\
a Gaussian spectrum centered at $\omega_p/2$:
\begin{align}\label{eq_gau2}
G(\omega_1,\omega_2)\approx&\sqrt{\frac{2}{\pi}}\frac{1}{\Delta\omega}\exp\left[-2\left(\frac{\omega_1-\omega_p/2}{\Delta\omega}\right)^2\right]\nonumber\\
&\times\delta(\omega_p-\omega_1-\omega_2),
\end{align}\\
a Lorentzian spectrum centered at $\omega_p/2$:
\begin{equation}\label{eq_lor2}
G(\omega_1,\omega_2)\approx\frac{2}{\pi}\frac{(\Delta\omega)^{3}\delta(\omega_p-\omega_1-\omega_2)}{\left[(\Delta\omega)^2+(\omega_1-\omega_P/2)^2\right]^2},
\end{equation}\\
a rectangular spectrum centered at $\omega_p/2$:
\begin{equation}\label{eq_rec}
G(\omega_1,\omega_2)\approx
\begin{cases}\frac{\delta(\omega_p-\omega_1-\omega_2)}{2\Delta \omega} & |\omega_1-\omega_p/2|\leqslant\Delta\omega\\ 0 &|\omega_1-\omega_p/2|>\Delta\omega
\end{cases},
\end{equation}\\
a double-Gaussian spectrum centered at
\mbox{$(\omega_p-5\Delta\omega)/2$} and
\mbox{$(\omega_p+5\Delta\omega)/2$}:
\begin{align}\label{eq_dougau2}
G(\omega_1,\omega_2)&\approx\frac{1}{\sqrt{2\pi}\Delta\omega}\delta(\omega_p-\omega_1-\omega_2)\nonumber\\
&\times\Bigg\{\exp\left[-2\left(\frac{\omega_1-(\omega_p-5\Delta\omega)/2}{\Delta\omega}\right)^2\right]\nonumber\\
&+\exp\left[-2\left(\frac{\omega_1-(\omega_p+5\Delta\omega)/2}{\Delta\omega}\right)^2\right]\Bigg\}.
\end{align}
a double-Lorentzian spectrum centered at
\mbox{$(\omega_p-30\Delta\omega)/2$} and
\mbox{$(\omega_p+30\Delta\omega)/2$}:
\begin{align}\label{eq_doulor2}
G(\omega_1,\omega_2)&\approx\frac{1}{\pi}(\Delta\omega)^3\delta(\omega_p-\omega_1-\omega_2)\nonumber\\
&\times\big\{\left[(\Delta\omega)^2+(\omega_1-\omega_p/2+15\Delta\omega)^2\right]^{-2}\nonumber\\
&+\left[(\Delta\omega)^2+(\omega_1-\omega_p/2-15\Delta\omega)^2\right]^{-2}\big\}.
\end{align}
Through the analogous analysises in Sec. \ref{section_single}, we
can obtain the correlation function $\mathcal{G}(l)$, the linear
entropy $S_L(l)$ and the concurrence $C(l)$, and investigate the
different decoherence processes. For simplicity, we plotted the
behaviors of linear entropy (see Fig.~\ref{fig_s2}) and concurrence
(see Fig.~\ref{fig_c2}) of the specific initial polarization state
$\left(\ket{HH}+\ket{VV}\right)/\sqrt{2}$ against the crystal length
$l$ in the case of the four FSEs shown above (again we set $\Delta
n\Delta \omega/c=500m^{-1}$).

We can see that coherence and concurrence vanish asymptotically in
the case of the Gaussian spectrum and the Lorentzian spectrum, while
vanish periodically with damped revival amplitudes in the case of
the rectangular spectrum, the double-Gaussian spectrum and the
double-Lorentzian spectrum.

\section{Conclusions and discussions}\label{sec_conclude}

We have considered the effects of photon spectrum forms on the
decoherence process in phase-damping channels. We have found that
the behavior of coherence depends on the choice of FSE. We have also
given some examples of the forms of FSE to show different behaviors.
An ideal white spectrum gives rise to an ideal Markovian decoherence
process, while, color spectrums result in non-Markovian correlation
functions, inducing non-Markovian effects on the decoherence
process. Explicitly, among our examples, against the crystal length,
coherence was found to vanish asymptotically in the case of a
Gaussian spectrum and a Lorentzian spectrum, while periodically with
variable revival amplitudes in the case of a rectangular spectrum
and a multi-peaked spectrum.

We would like to discuss the experimental feasibility of our results
briefly. In our paper, we only gave theoretical analysises with
several ideal choices of FSE, however, analogous analysises can
appply to any practical spectrum forms. Furthermore, photons with
various FSEs have been experimentally realized. The spectrum filter
usually used may restrict the FSE to a Gaussian form
\cite{Berglund_decoherence}. Quantum dot \cite{source_moreau,
source_Santori} and fluorescence \cite{source_Brunel, source_Lounis,
source_wrigge, source_vamivakas} based sources may generate photons
with a Lorentzian spectrum. Keller \emph{et al.}
\cite{source_Keller} have demonstrated the productions of Gaussian,
rectangular, and double-peaked wave functions of photons emitted
from Raman pumped single ions trapped in a cavity by manipulating
the pulse pump. The case of Double-Lorentzian functions may find
possible applications to the case of photonic bandgap. There have
been a few experimental reports on shaping wave packets of entangled
photons from parametric down-conversion \cite{shaping_bellini,
shaping_viciani, shaping_peer, shaping_baek}. Moreover, it should be
noted that some of us have ever observed the coherence revivals in
the phase-damping channels by restricting the photon spectrum using
filters with a rectangular transmission function (see Appendix in
Ref.~\cite{rec_zhang}), and that during our preparation of this
paper, we became aware that, by shaping the spectrum of photons with
a Fabry-Perot Cavity, coherence revivals of single- and two-photon
polarization states in the phase-damping channels had been observed
\cite{xu2008}.

Besides the Refs.~\cite{shaping_bellini, shaping_viciani,
shaping_peer, shaping_baek}, the spectral influences on the photon
correlations or fourth-order interference \cite{spectrum_rubin,
spectrum_Keller, spectrum_Grice, spectrum_Grice2, spectrum_andrews,
spectrum_rohde} have been extensively studied. Recently, the
decoherence of photon pairs from parametric down-conversion have
been experimentally studied and characterized in terms of
fluctuations of pump laser \cite{source_Cialdi, source_Cialdi2}. Our
approach would be possible to find applications with these studies.
We hope our work can stimulate more investigations on the
characteristics of photon spectrum distributions. For instance,
further results beyond the studies on polarization mode dispersion
in optical fibers \cite{PMD_zhang, PMD_poon} would be obtained if
different types of FSE could be considered.

\begin{acknowledgments}
Y.X.G. thanks Chuan-Feng Li, Jin-Shi Xu and Chao-Yang Lu for helpful
discussions. This work was funded by National Fundamental Research
Program (Grant No. 2006CB921907), National Natural Science
Foundation of China (Grants No. 10674127, No. 60621064 and No.
10774139), Innovation Funds from Chinese Academy of Sciences,
Program for New Century Excellent Talents in University, A
Foundation for the Author of National Excellent Doctoral
Dissertation of PR China (grant 200729).
\end{acknowledgments}

\bibliography{dephasingref}
\end{document}